\documentclass[useAMS,usenatbib]{mn2e}

\def\HI{\ifmmode{\rm HI}\else{H\/{\sc i}}\fi}

\def\lsun{\ifmmode{{\mathrm L}_{\odot}}\else{L$_{\odot}$}\fi}

\def\msun{\ifmmode{{\mathrm M}_{\odot}}\else{M$_{\odot}$}\fi} 
\def\msunpc2{\ifmmode{{\mathrm M}_{\odot} \, {\mathrm{pc}}^{-2}}\else{M$_{\odot} \, {\mathrm {pc}}^{-2}$}\fi}

\def\kms{\ifmmode{{\mathrm{km \, s^{-1}}}}\else{${\mathrm{km \, s^{-1}}}$}\fi}

\def\la{\mathrel{\mathchoice {\vcenter{\offinterlineskip\halign{\hfil
$\displaystyle##$\hfil\cr<\cr\sim\cr}}}
{\vcenter{\offinterlineskip\halign{\hfil$\textstyle##$\hfil\cr
<\cr\sim\cr}}}
{\vcenter{\offinterlineskip\halign{\hfil$\scriptstyle##$\hfil\cr
<\cr\sim\cr}}}
{\vcenter{\offinterlineskip\halign{\hfil$\scriptscriptstyle##$\hfil\cr
<\cr\sim\cr}}}}}
\def\ga{\mathrel{\mathchoice {\vcenter{\offinterlineskip\halign{\hfil
$\displaystyle##$\hfil\cr>\cr\sim\cr}}}
{\vcenter{\offinterlineskip\halign{\hfil$\textstyle##$\hfil\cr
>\cr\sim\cr}}}
{\vcenter{\offinterlineskip\halign{\hfil$\scriptstyle##$\hfil\cr
>\cr\sim\cr}}}
{\vcenter{\offinterlineskip\halign{\hfil$\scriptscriptstyle##$\hfil\cr
>\cr\sim\cr}}}}}

\usepackage{times} 
\usepackage{graphicx}
\usepackage{rotating}
\usepackage{epsfig}
\usepackage{amsmath}

\def\funits{M_\odot{\rm kpc}^{-3}({\rm km s}^{-1})^{-3}\,}
\def\fiestas{FiEstAS\,}
\def\enbid{EnBiD\,}
\def\beq{\begin{equation}}
\def\eeq{\end{equation}}

\title[Evolution of phase-space density in LCDM halos]{Evolution of the Dark
Matter Phase-Space Density Distributions of $\Lambda$CDM Halos}

\author[Vass~et~al.]{Ileana M. Vass$^{1,2}$, Monica
Valluri$^{2,3}\thanks{direct queries to: mvalluri@umich.edu (MV)}$, Andrey V.
Kravtsov$^{2,4,5}$, Stelios Kazantzidis$^{6}$ \\
$^{1}$ Department of Astronomy University of Florida, Gainesville, FL 32611,
USA \\
$^{2}$ Kavli Institute for Cosmological Physics, The University of Chicago,
Chicago, IL 60637, USA,  \\
$^{3}$ Department of Astronomy, University of Michigan,  Ann Arbor, MI 48109,
USA, {\tt mvalluri@umich.edu}\\
$^{4}$ Enrico Fermi Institute, The University of Chicago, Chicago, IL 60637,
USA \\
$^{5}$ Dept. of Astronomy \& Astrophysics, The University of Chicago, 5640 S.
Ellis Ave., Chicago, IL 60605\\
$^{6}$ Center for Cosmology and Astro-Particle Physics, The Ohio State
University, Columbus, OH 43210, USA}
\begin{document}

\date{Accepted 2009 Feburary 4; Received 2009 February 2; in original form 2008 October 1}

\pagerange{\pageref{1225}--\pageref{1236}} \pubyear{2009w}

\maketitle

\label{firstpage}
\begin{abstract}  

We study the evolution of phase-space density during the hierarchical
structure formation of $\Lambda$CDM halos.  We compute both a
spherically-averaged surrogate for phase-space density ($Q =
\rho/\sigma^3$) and the coarse-grained distribution function
$f(\mathbf{x},\mathbf{v})$ for dark matter particles that lie within
$\sim 2$ virial radii of four Milky-Way-sized dark matter halos. The
estimated $f(\mathbf{x}, \mathbf{v})$ spans over four decades at any
radius. Dark matter particles that end up within two virial radii of a
Milky-Way-sized DM halo at $z=0$ have an approximately Gaussian
distribution in $\log(f)$ at early redshifts, but the distribution
becomes increasingly skewed at lower redshifts. The value $f_{\rm
peak}$ corresponding to the peak of the Gaussian decreases as the
evolution progresses and is well described by $f_{\rm peak}(z) \propto
(1+z)^{4.5}$ for $z > 1$.   The highest values of $f$, (responsible for the skewness of the profile)
are found at the centers of dark matter halos and subhalos, where $f$
can be an order of magnitude higher than in the center of the main
halo.  We confirm that $Q(r)$ can be described by a power-law with a
slope of $-1.8 \pm 0.1$ over 2.5 orders of magnitude in radius
and over a wide range of redshifts. This $Q(r)$ profile likely
reflects the distribution of entropy ($K \equiv \sigma^2/\rho_{\rm
dm}^{2/3} \propto r^{1.2}$), which dark matter acquires as it is
accreted onto a growing halo. The estimated $f(\mathbf{x},
\mathbf{v})$, on the other hand, exhibits a more complicated
behavior. Although the median coarse-grained phase-space density
profile $F(r)$ can be approximated by a power-law, $\propto r^{-1.6
\pm 0.15}$, in the inner regions of halos ($<0.6 \; r_{\rm vir}$), at
larger radii the profile flattens significantly. This is because
phase-space density averaged on small scales is sensitive to the
high-$f$ material associated with surviving subhalos, as well as
relatively unmixed material (probably in streams) resulting from
disrupted subhalos, which contribute a sizable fraction of matter at
large radii.

\end{abstract}


\bigskip
}
\section{Introduction}
\label{sec:introduction}

Cosmological $N$-body simulations of the formation of structure in the
Universe allow us to reconstruct how gravitationally bound objects
like galaxies and clusters form and evolve. These simulations have
shown that despite the seemingly complex hierarchical formation
history of dark matter (DM) halos, the matter density distributions of
cosmological DM halos are described by simple 3-parameter profiles
\citep{NFW, Metal2006}.

The study of the evolution of the phase-space distribution function
(DF) of a collisionless dynamical system (composed of stars or DM or
both) is fundamental to understanding its dynamical evolution, since a
collisionless system is completely described by its phase-space
density distribution (otherwise called the fine-grained DF
$\mathbf{f}(\mathbf{x}, \mathbf{v})$). For an isolated collisionless
system, $\mathbf{f}(\mathbf{x}, \mathbf{v}, t)$ is fully described by
the Collisionless Boltzmann Equation (CBE) (sometimes called the
Vlasov equation), which is a continuity equation in phase-space. A
consequence of the CBE is that the fine-grained DF is always
conserved.  In addition, $\mathbf{V}(\mathbf{f})d\mathbf{f}$, the
volume of phase-space occupied by phase-space elements whose density
lies between $(\mathbf{f},\mathbf{f}+d\mathbf{f})$, is also conserved.
However, the conservation of the fine-grained $\mathbf{f}(\mathbf{x},
\mathbf{v})$ and $\mathbf{V}(\mathbf{f})d\mathbf{f}$ is not a very
useful property for understanding the evolution of real systems, since
in practice it is only possible to compute an average of $\mathbf{f}$
over a finite volume of phase-space. This average is referred to as
the coarse-grained DF $f(\mathbf{x}, \mathbf{v})$, and the associated
volume density is $V(f)$.  The evolution of the coarse-grained DF is
governed by {\em Mixing Theorems} \citep{BT,
tremaine_henon_lyndenbell_86, mathur_88}. These theorems state that if
the coarse-grained DF is bounded, then processes that operate during
the relaxation of collisionless systems (e.g.  phase mixing, chaotic
mixing, and the mixing of energy and angular momentum that accompanies
violent relaxation) result in a decrease in the coarse-grained
phase-space density. This is because at any point in phase-space, both
low and high phase-space density regions get highly mixed \citep{BT,
tremaine_henon_lyndenbell_86}.  In addition, there are simple
relationships that exist between the initial $\mathbf{V}(\mathbf{f})$
and the coarse-grained $V(f)$ at a later time: $V(f)$ is always
greater than $\mathbf{V}(\mathbf{f})$ \citep{mathur_88}.

Recently, \citet{dehnen2005} proved a new Mixing Theorem
that states that the excess-mass function $D(f)$, which is the mass of
material with coarse-grained phase-space density greater than a value
$f$, always decreases due to mixing, for all values of $f$.  He
showed, using this theorem, that steeper cusps are less mixed than
shallower ones, independent of the details of the DF or density
profile, and this implies that a merger remnant cannot have a cusp
that is either steeper or shallower than the steepest of its
progenitors.

\citet{taylorN01} have introduced a surrogate quantity, $Q(r) =
\rho(r)/\sigma(r)^{3}$, where $\rho(r)$ is the configuration-space
density averaged in spherical shells, and $\sigma(r)$ is the velocity
dispersion of DM particles averaged in a spherical shell
centered at the radius $r$.  Note that, although $Q(r)$ has dimensions
of phase-space density, it is  {\it not} true coarse-grained
phase-space density, because it is constructed out of separately
computed configuration-space density and velocity dispersion in
arbitrarily chosen spherical shells. Consequently, it is not expected
to obey any {\em Mixing Theorems} \citep{dehnen_mclaughlin_05} and is
generally difficult to interpret in terms of phase-space
density. Nevertheless, it is an intriguing quantity, because
\citet{taylorN01} show that $Q(r)$ is well approximated by a single
power-law, $Q \propto r^{-\beta} $ with $\beta \approx 1.874$ over
more than 2.5 orders of magnitude in radius for CDM halos universally,
regardless of mass and background cosmology \citep[see
also][]{rasia_etal04,ascasibar_etal04}. Recent work, however,
indicates that $Q(r)$ profiles are somewhat sensitive to the amount of
substructure and the slope of the power spectrum
\citep{wang_white08,knollman_etal08}.

While power-law $Q(r)$ profiles are by no means universal for
self-gravitating systems in dynamical equilibrium \citep{barnes06},
power-law profiles with exactly the same slope were first shown to
arise in simple self-similar spherical gravitational infall models
\citep{bertschinger_85} pointing to a possible universality in the
mechanisms that produces them. \citet{austin05} extended the early
work of \citet{bertschinger_85} using semi-analytical extended
secondary infall models to follow the evolution of collisionless
spherical shells of matter that are initially set to be out of
dynamical equilibrium and are allowed to move only radially.  They
concluded that the power-law behavior of the final phase-space density
profile is a robust feature of virialized halos, which have reached
equilibrium via violent relaxation. They used a constrained Jeans
equation analysis to show that this equation has different types of
solutions and that it admits a unique periodic solution which gives a
power-law density profile with slope $\beta = 1.9444$ (comparable to
the numerical value of $\beta = 1.87$ obtained by \citet{taylorN01}).

A recent study by \citet{hoffman07} followed the evolution of $Q(r)$
profiles in cosmological DM halos in constrained simulations designed
to control the merging history of a given halo. These authors showed
that, during relatively quiescent phases of halo evolution, the
density profile closely follows that of a NFW halo, and $Q(r)$ is
always well represented by a power-law $r^{-\beta}$ with $\beta =
1.9\pm 0.1$. They showed that $Q(r)$ deviates from power-law most
strongly during major mergers but recovers the power-law form
thereafter \citep[this is consistent with our findings using a series
of controlled merger simulations, ][]{vass_etal_08a}.  More recently,
\citet{wojtak_etal08} demonstrated (using a variant of the \fiestas
code) that the DF of $\Lambda$CDM halos of mass $10^{14}-
10^{15}~M_{\odot}$ can be separated into energy and angular momentum
components. They proposed a phenomenological model for spherical
potentials and showed that their model DF was a good match to the
N-body DFs and was also able to reproduce the power-law behavior of $Q(r)$.  Another
study of relevance is the work of \citet{pei_fr_pac07}, which defined
a global value for the phase-space density, $Q$, which they found
decreases rapidly with time. Despite the insights obtained in these
studies, the origin for  the universality of $Q(r)$ is not yet
understood, and the question of how this quantity relates to the true
coarse-grained phase-space density has not yet been thoroughly
investigated.  This motivates our study which is focused on the
evolution of {\it both} $Q(r)$ and phase-space density evolution using
a set of self-consistent cosmological simulations of halo formation.

In the last four years, three independent numerical codes to compute
the true coarse-grained phase-space density from $N$-body simulations
have been developed \citep{Arad04, ascasibar_binney05, shar_stein06}.
In the Appendix we compare results obtained with the codes
``\fiestas'' \citep{ascasibar_binney05} and ``\enbid''
\citep{shar_stein06} at a range of redshifts. However we only present results obtained with
the ``\enbid'' code using a specific kernel.  Following the submission of this paper we became
aware of the work of \citet{maciejewski_etal_08}. These authors carried out a
similar comparison of various  coarse-grained phase
space density estimators for N-body simulations  cosmological halos at $z=0$.

In an accompanying paper \citep{vass_etal_08a}, we present analysis of
phase-space and $Q(r)$ evolution during major mergers between CDM-like
spherical halos.  We show that major mergers (those which are the most
violent and therefore likely to result in the greatest amount of
mixing in phase-space) preserve spherically- averaged phase-space
density profiles (both $Q(r)$ and $F(r)$) out to the virial radius
$r_{\rm vir}$ of the final remnant.  We confirm the prediction
\citep{dehnen2005} that the phase-space density profiles of DM halos
are extremely robust, and in particular, the prediction that the
steepest central cusp always survives.

To understand the robustness of profiles outside the central cusp
\citep{vass_etal_08a}, we drew on recent work by \citet{valluri07}
that shows that matter in equally-spaced radial shells is
redistributed during the merger is such a way that only about 20\% of
the matter in the central cusp is ejected during the merger to radii
extending out to about two scale radii. For all other radii, roughly
15\% by mass in each radial shell is redistributed uniformly with
radius.  Further, nearly 40\% of the mass of the progenitor halos lies
beyond the formal virial radius of the remnant
\citep{kazantzidis_etal06, valluri07}, and this matter originates
roughly uniformly from each radial interval starting from about three
scale radii from the center.  While major mergers \citep[such as those
studied by][]{dehnen2005,vass_etal_08a} are instructive, they
represent the most extreme form of mixing, and are not the major mode
of mass accretion in the Universe. In addition, since experiments with
major mergers demonstrate that pre-existing power-law profiles are
preserved, they shed little light on the formation of the
profiles. Our primary goal is to gain a better understanding of the
origin of the power-law phase-space density profiles seen in
cosmological DM simulations.

In this paper we investigate the evolution of the coarse-grained
phase-space density in the formation and evolution of four
Milky-Way-sized halos in a $\Lambda$CDM cosmology with cosmological
parameters:
$(\Omega_{m},\Omega_{\Lambda},h,\sigma_{8})=(0.3,0.7,0.7,0.9)$.  In
\S~\ref{sec:numerical} we summarize the cosmological N-body
simulations analyzed in this study as well as the numerical methods
used to obtain coarse-grained phase-space densities. (The Appendix
contains a detailed comparison of the two codes \fiestas and
\enbid. We also present reasons for our choice of \enbid, with a
$n=10$ smoothing kernel).  In \S~\ref{sec:onehalo} we describe the
properties of the spherically-averaged quantity $Q(r)$ as well as the
the coarse-grained DF for one Milky-Way-sized DM halo at $z=0$, as
well as the evolution of the phase-space DF of this halo from $z=9$ to
the present. In \S~\ref{sec:otherhalos} we also present results for
three additional Milky-Way-sized halos. In \S~\ref{sec:origin} we
present an interpretation of the power-law profiles seen in $Q(r)$ and
discuss the implications of observed coarse grained distribution
function in a cosmological context. \S~\ref{sec:conclude} summarizes
the results of this paper and concludes.

\bigskip
\section{Numerical  Methods}
\label{sec:numerical}

The simulations analyzed in this paper are described in greater detail
in the works of \citet{tumultuos} and \citet{fossils}. The simulations
were carried out using the Adaptive Refinement Tree $N$-body code
\citep[ART][]{art}. The simulation starts with a uniform $256^{3}$
grid covering the entire computational box.  This grid defines the
lowest (zeroth) level of resolution. Higher force resolution is
achieved in the regions corresponding to collapsing structures by
recursive refining of all such regions by using an adaptive refinement
algorithm. Each cell can be refined or de-refined individually. The
cells are refined if the particle mass contained within them exceeds a
certain specified value. The grid is thus refined to follow the
collapsing objects in a quasi-Lagrangian fashion.

Three of the galactic DM halos were simulated in a comoving box of 25
$h^{-1}$ Mpc (hereafter $L25$); they were selected to reside in a
well-defined filament at $z = 0$. Two halos are neighbors, located 425
$h^{-1}$ kpc from each other.  The third halo is isolated and is
located two Mpc away from the pair.  Hereafter, we refer to the isolated
halo as $G_1$ and the halos in the pair as $G_2$ and $G_3$.  Although we analyzed all three halos at the same level of detail
we present detailed results for $G_1$.  The virial masses
and virial radii for the halos studied are given in Table 1 in
\citet{tumultuos}. The virial radius (and the corresponding virial
mass) was chosen as the radius encompassing a mean density of $\sim 200$
times the mean density of the Universe.  The masses of the DM halos
are well within the range of possibilities allowed by models for the
halo of the Milky Way galaxy \citep{klypin02}.  The simulations
followed a Lagrangian region corresponding to the sphere of radius
equal to two virial radii around each halo at $z=0$.  This region was
re-sampled with the highest resolution particles of mass $m_{\rm
p}=1.2\times 10^6h^{-1}{\rm M_{\odot}}$, corresponding to $1024^3$
particles in the box, at the initial redshift of the simulation
($z_{\rm i}=50$).  The maximum of ten refinement levels was reached in
the simulations corresponding to the peak formal spatial resolution of
$152h^{-1}$ parsec.  Each host halo is resolved with $\sim 10^6$
particles within its virial radius at $z=0$.

The fourth halo was simulated in a comoving box of 20 $h^{-1}$ Mpc box
(hereafter $L20$), and it was used to follow the Lagrangian region of
approximately five virial radii around the Milky Way-sized halo with
high resolution.  In the high-resolution region the mass of the DM particles is $m_{\rm p}=6.1\times 10^5h^{-1}{\,\rm}M_{\odot}$,
corresponding to effective $1024^3$ particles in the box, at the
initial redshift of the simulation ($z_{\rm i}=70$).  As in the other
simulation, this run starts with a uniform $256^3$ grid covering the
entire computational box. Higher force resolution is achieved in the
regions corresponding to collapsing structures by recursive refining
of all such regions using an adaptive refinement algorithm. Only
regions containing highest resolution particles were adaptively
refined.  The maximum of nine refinement levels was reached in the
simulation corresponding to the peak formal spatial resolution of
$150h^{-1}$ comoving parsec.  The Milky Way-sized host halo has the
virial mass of $1.4\times 10^{12}h^{-1}\,\rm M_{\odot}$ (or $2.3$
million particles within the virial radius) and virial radius of
$230h^{-1}{\rm kpc}$ at $z=0$.

The peak spatial resolution of the simulations determines the minimum
radius to which we can trust the density and velocity profiles. In the
following analysis, we only consider profiles at radii at least eight
times larger than the peak resolution of the simulations (i.e.,
$\approx 0.5-0.6h^{-1}$~comoving kpc or $r\approx 0.004\, r_{\rm
vir{(z=0)}}$).

\bigskip
\subsection{Phase-space density estimators}
\label{sec:estimators}

In this paper, we will focus on the time evolution of the spherically
averaged phase-space density $Q(r)$ as well as the coarse-grained
phase-space DF $f(\mathbf{x}, \mathbf{v})$ during the formation of the
Milky-Way sized DM halos in $\Lambda$CDM cosmological simulations
described above.

In the last four years three independent numerical codes to compute
the coarse-grained phase-space density from $N$-body simulations have
been developed \citep{Arad04, ascasibar_binney05,
shar_stein06}\footnote{An even more sophisticated numerical method has
recently been presented by \citet{vogelsberger_etal08} for calculating
the fine-grained phase-space structure of DM distributions
derived from cosmological simulations. This code has the potential to
identify fine-scale structure such as caustics in phase-space and the
phase-space structure of tidal streams in the Milky-Way halo. Its
ability to estimate the fine-grained $\mathbf{f}(\mathbf{x},
\mathbf{v})$ makes it useful for understanding non-equilibrium systems
and for making accurate predictions for direct DM searches.}. These
techniques differ in the scheme used to tesselate 6-dimensional phase
space as well as in the density estimators they use.  The first study
of coarse grained DF \citep{Arad04} showed that it has its highest
values at the centers of DM halos and subhalos. They also showed that
the volume density of phase-space $V(f)$ within individual
cosmological DM halos at $z=0$ has a power-law profile over
nearly four orders of magnitude in $f$. The main criticism of this
code is that it is not metric-free.  The FiEstAS algorithm of
\citet{ascasibar_binney05} and the EnBid algorithm of
\citet{shar_stein06} are preferred due to their speed and the metric
free nature.  The {\fiestas} algorithm is based on a repeated division
of each dimension of phase-space into two regions that contain roughly
equal numbers of particles.  \enbid \citep{shar_stein06} closely
follows the method used in \fiestas, but it optimizes the number of
divisions to be made in a particular dimension by using a
minimum-entropy criterion based on the {\it Shannon entropy} that
measures the phase-space density with much greater accuracy when $f$
is high.  \citet{shar_stein06} showed that \enbid with a kernel which
includes 10 nearest neighbors ($n=10$) about each point was able to
recover analytic phase-space density profiles to nearly 3-4 decades
higher values of $f$ than \fiestas.

For our halos at $z=0$, the results we obtained with the different
codes were completely consistent with those obtained by
\citet{shar_stein06}. The deviations between the estimates obtained
with the different codes and various parameters were significantly
higher at higher redshift - where comparisons with analytic estimates
are not available. We refer the reader to the Appendix for a detailed
comparison between estimated values of $f$ using the two codes at
various redshifts. In the rest of this paper we present results
obtained with \enbid ($n=10$ kernel), the parameters preferred by
\citet{shar_stein06}.

\begin{figure}
\centerline{\psfig{figure=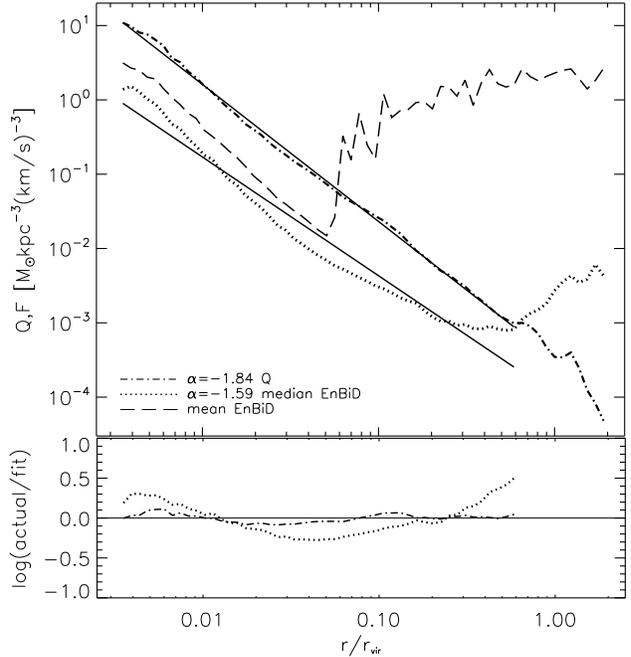,width=8.5cm}}
\caption{$Q(r)$ and $F(r)$ in 100 spherical radial bins about the
most-bound-particle in halo $G_1$ from the $L25$ simulation at $z$=0.  Top
panel: $Q$ (dot-dashed curve); power-law fit $Q \propto r^{-1.84 \pm
0.012}$ is given by upper thin solid line.  $F(r)$ from \enbid \, (with $n=10$
smoothing kernel (dotted curve)); power-law fit to $F(r) \propto r^{-1.59 \pm
0.054}$ is given by lower thin solid line.  Mean value of
$\mathbf{f}$ (long-dashed curve) is noisier  than $F(r)$ due to subhalos. 
Residuals to the two power-law
fits are shown in the bottom panel.
\label{fig:qandfz0}}
\end{figure}

Figure~\ref{fig:qandfz0} {\it Top} shows the spherically averaged
quantity $Q =\rho/\sigma^{3}$ (dot-dashed curve); the mean of
$f(\mathbf{x}, \mathbf{v}$) in spherical bins (dashed curve) and the
median DF in spherical bins (dotted curve).  As we will show in
Figure~\ref{fig:rf_L25_cont}, the large fluctuations in the mean value
of $f$ (dashed curve) beyond $0.1r_{\rm vir}$ are due to the presence
of substructure, which can have extremely high central values of $f$.
The median value of $f$ (hereafter represented by $F$) is much
smoother, being less sensitive to the large range (nearly eight orders
of magnitude) in $f$ at each radius.  In what follows, we will use the
median $F$, computed in concentric radial bins centered on the most
bound particle in the main halo, since it is less sensitive to
substructure.  $Q(r)$ is well fitted by a power-law: $Q \propto
r^{-1.84\pm 0.012}$ over the radial range $[0.004$--$0.6] \, r/r_{\rm
vir}$ (upper thin solid line).  The power-law fit $F(r) \propto
r^{-1.59 \pm 0.054}$ over the same radial range is shown by the lower
thin solid line. The bottom panel shows the residuals of the fits $F$
($\log(F/F_{\rm fit})$) and $Q$ ($\log(Q/Q_{\rm fit})$). This plot
shows that while $Q(r)$ is an extremely good power-law, the median
$F(r)$ is only approximately power-law over the same radial range. Note that
$Q(r)$ is numerically larger than $F(r)$ due to the fact that the former quantity does not properly account for the volume of the phase-space element. A
similar result was obtained in a much higher resolution simulation by
\citet{stadel_etal08}, a study which appeared as we were preparing
this paper for publication.

\bigskip
\section{Evolution of phase-space density with redshift}
\label{sec:onehalo}

\citet{hoffman07} studied the phase-space density profiles of a DM halo by tracking $Q(r)$ for material within the formal virial
radius at each redshift.  They found that the virialized material
within this radius has an approximately power-law form with a constant
power-law index of $-1.9\pm 0.05$ at all redshifts from $z=5$ to the present.

We follow a slightly different approach in this paper, since our
objective is to understand the evolution of the true coarse-grained
phase-space density distribution. We track the evolution of
phase-space density, by tracing backwards in time, all the material
that lies inside the virial radius at $z=0$. This will allow us to
understand how the initially high phase-space density material that
lies outside virialized systems at high redshift falls in and
undergoes mixing and how the resultant mixing preserves the
phase-space density profiles as a function of redshift.

We identified Milky-Way sized DM halos at $z=0$ in our cosmological
simulations and identify all the particles that lie inside twice the
virial radius at $z=0$. Our choice of halo outer radius (2$r_{\rm
vir}$) is motivated by the recent work of \citet{cuesta_etal_08} who
showed that the virial radius is arbitrary and does not correspond to
a physically meaningful outer boundary for a DM halo. A better {\it
outer} radius is the so-called ``static radius'', defined as the
radius at which the mean radial velocity of particles is zero. This
radius is typically about $2r_{\rm vir}$ for a galaxy-sized DM halo.
After identifying all particles within  $2r_{\rm vir}$ at $z=0$, we
tracked them back to $z=9$. Our simulations do proceed back in time to
even higher redshifts, however we do not analyze them here because the
mass resolution at higher redshifts does not allow us to resolve the
evolution of most of the objects beyond this epoch.  There were
1.6$\times 10^{6}$ particles within 2 virial radii of the $G_1$ halo,
which is the halo that is most isolated at $z=0$. Two other halos
($G_2$ and $G_3$) from the $L25$ simulation are closer to each other
at $z=0$ (1.5 virial radii apart).  Therefore, we only tracked those
particles which lay within one virial radius of the centers of these
two halos at $z=0$ back in time to $z=9$.  The particles in the high
resolution simulation $L20$ were selected and tracked in the same way
as for the $G_1$ halo, and at $z=0$ there were 3$\times 10^{6}$
particles inside twice the virial radius.  The position and velocity
data for all the particles identified as belonging to a given halo at
$z=0$ were analysed.

 At all redshifts we compute the phase-space densities of
particles in {\em physical coordinates} rather than in co-moving
coordinates.  Distances are always in kpc and velocities are in $\rm
{km {s}^{-1}}$, and these quantities are measured relative to the
center of the most massive progenitor. In many of the figures that
follow, the physical distances at all redshifts other than $z=0$ are
scaled by the virial radius at $z=0$; i.e. all distances are given in
units of $r/r_{\rm vir (z=0)}$.

In this section, we will present the results obtained
for the $G_1$ halo in the $L25$ simulation. A comparison with results for
the other three halos is made in \S~\ref{sec:otherhalos}.

\begin{figure}
\centerline{\psfig{figure=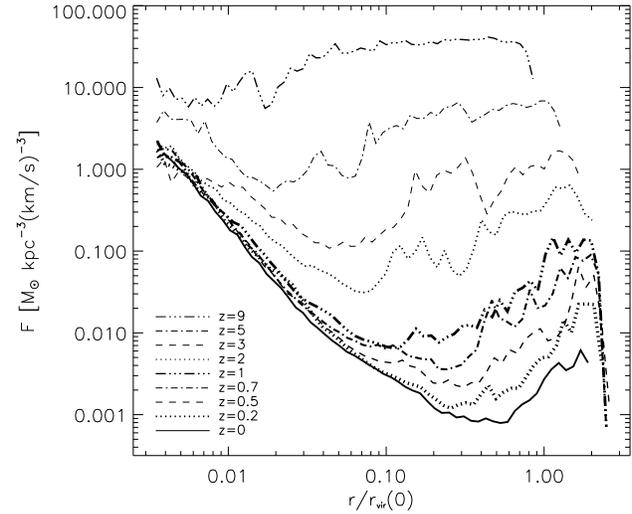,width=8.5cm}}
\caption{$F$ as a function of comoving radius $r$ for all material
that lies within 2$r_{\rm vir}$ of the center of the $G_1$
halo at $z=0$. Each curve corresponds to a different
redshift as indicated in the legends.}
\label{fig:f_allz}
\end{figure}

Figure~\ref{fig:f_allz} plots the median value ($F$) of the phase
space density $f$ as a function of radius (as defined above) at
several different redshifts.  At the highest redshift plotted ($z=9$),
the overall value of $F$ is higher at large radii than it is at the
center of the halo, and it only varies by a factor of few over the
entire range of radii ($F \sim 10-30\funits$).  This is because within
the inner, virialized regions the coarse-grained phase-space density
is lowered due to mixing, while it is still high in the outer regions
where the large fraction of matter is not yet mixed or is located in
small subhalos, that have not mixed to the same degree as the main
host.  As the evolution progresses, there is a decrease in the central
value of $F$ till $z= 3$.  The median value of $F$ at the center of
the halo drops from $F \approx 10\funits$ at $z=9$ down to $ F \approx
2\funits$ at $z=0$. The central value of $F$ remains constant beyond
$z=3$, while there is a steady decrease in $F$ at intermediate radii.

Although it is instructive to plot the mean and/or median values of
$F$ at each radius, much more information is contained in the full
coarse-grained phase-space density $f$.  The six-dimensional function
$f(\mathbf{x}, \mathbf{v})$ can be most easily visualized using the
volume DF $V(f)$, which also obeys a Mixing Theorem \citep{mathur_88}.
\citet{Arad04} showed that at $z=0$, $V(f)$ for cosmological DM halos
follows a power-law profile $V(f) \propto f^{-\alpha}$ with power-law
index $\alpha = 2.4$ over four orders of magnitude in $f$. They argued
that since DM halos are almost spherical, if their phase-space DFs are
approximately isotropic, their DFs could be written as functions of
energy alone: $f = f(E)$. In this case they showed that if $Q \propto
r^{-\beta}$, then $V(f)$ would also be described by a power-law, and
the index $\alpha$ was related to the index $\beta$ through a simple
equation.

\begin{figure*}
\centerline{\psfig{figure=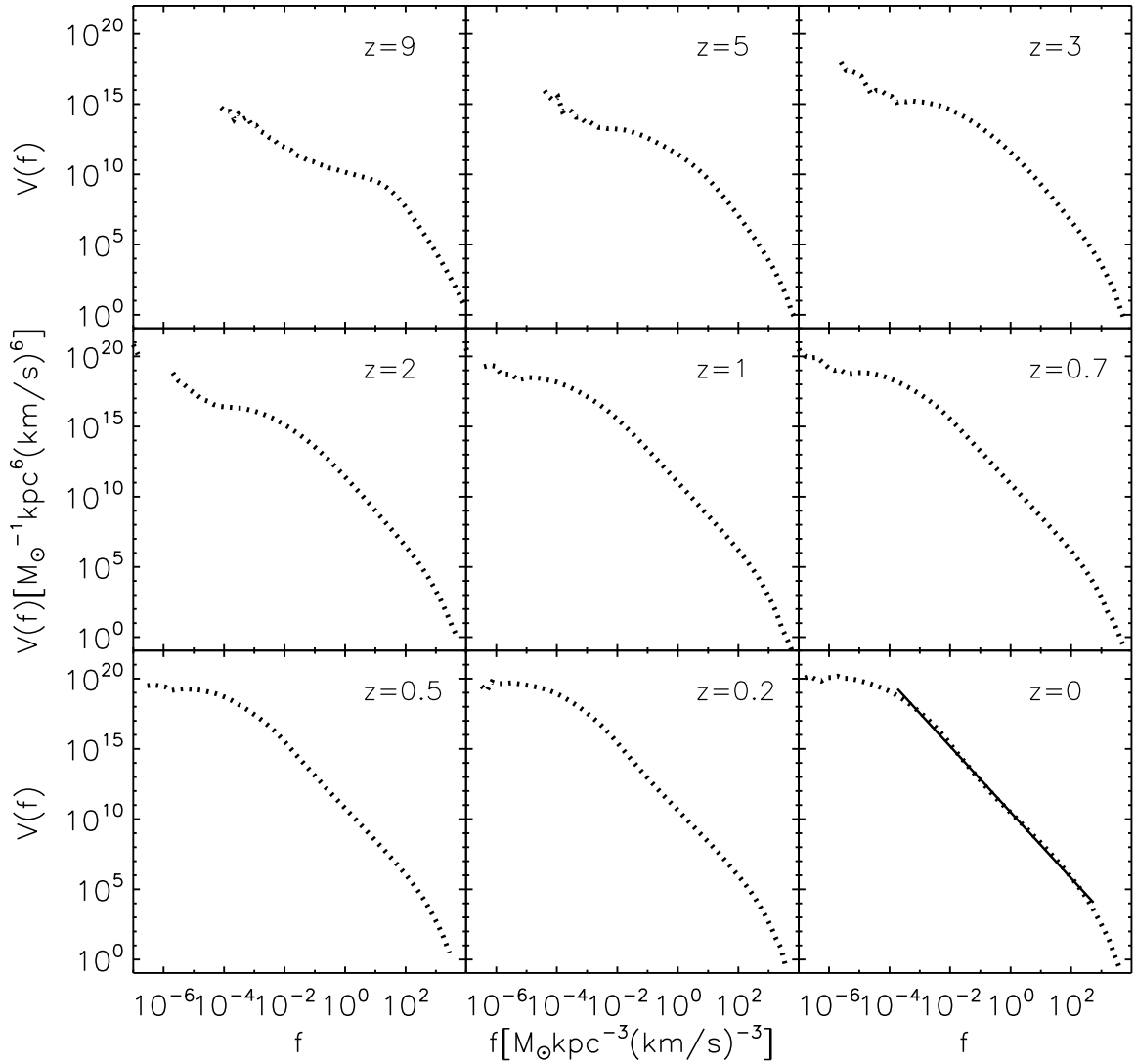,width=16.cm}}
\caption{The volume DF $V(f)$ for $G_1$ halo in the $L25$  simulation at different redshifts.  A power-law fit to $V(f) \propto f^{-2.34 \pm 0.017}$, is shown at $z=0$ (solid line).
\label{fig:vf_L25}}
\end{figure*}

In Figure~\ref{fig:vf_L25} we plot the evolution of $V(f)$ with
redshift for all the material that lies within twice the
virial radius of halo $G_1$ at $z=0$. 
In the bottom right-hand panel, the solid line is a fit to $V(f)$ for values of
$10^{-4} < f < 10^{2.5}$. Over this range the best fit power-law profile is
given by $V(f) \propto f^{-2.34 \pm 0.02}$ which is not very different from the power-law
profile fit obtained by \citet{Arad04}.  At all redshifts $V(f)$ deviates from
the power-law profile at both low and high end. It is likely that at the
high-$f$ end the distribution deviates from the power-law due to the unresolved
subhalos below the mass resolution limit of the simulation.

The {\it Mixing Theorem} requires that the coarse-grained volume
distribution function $V(f)$ always {\it increases} \citep{mathur_88}.
We see that  $V(f)$ increases  steadily from $z=9$
onward for $f < 1\funits$ in such a way that as the system evolves, there is more volume associated with
material with low $f$. At higher values of $f$, $V(f)$ remains almost constant
with decreasing redshift, probably owing to the fact that this material forms the cusps of DM halos.  While the Mixing Theorems have previously been
demonstrated for isolated systems, this is to our knowledge the first
demonstration of their validity in a cosmological context.

\begin{figure*}
\centerline{\psfig{figure=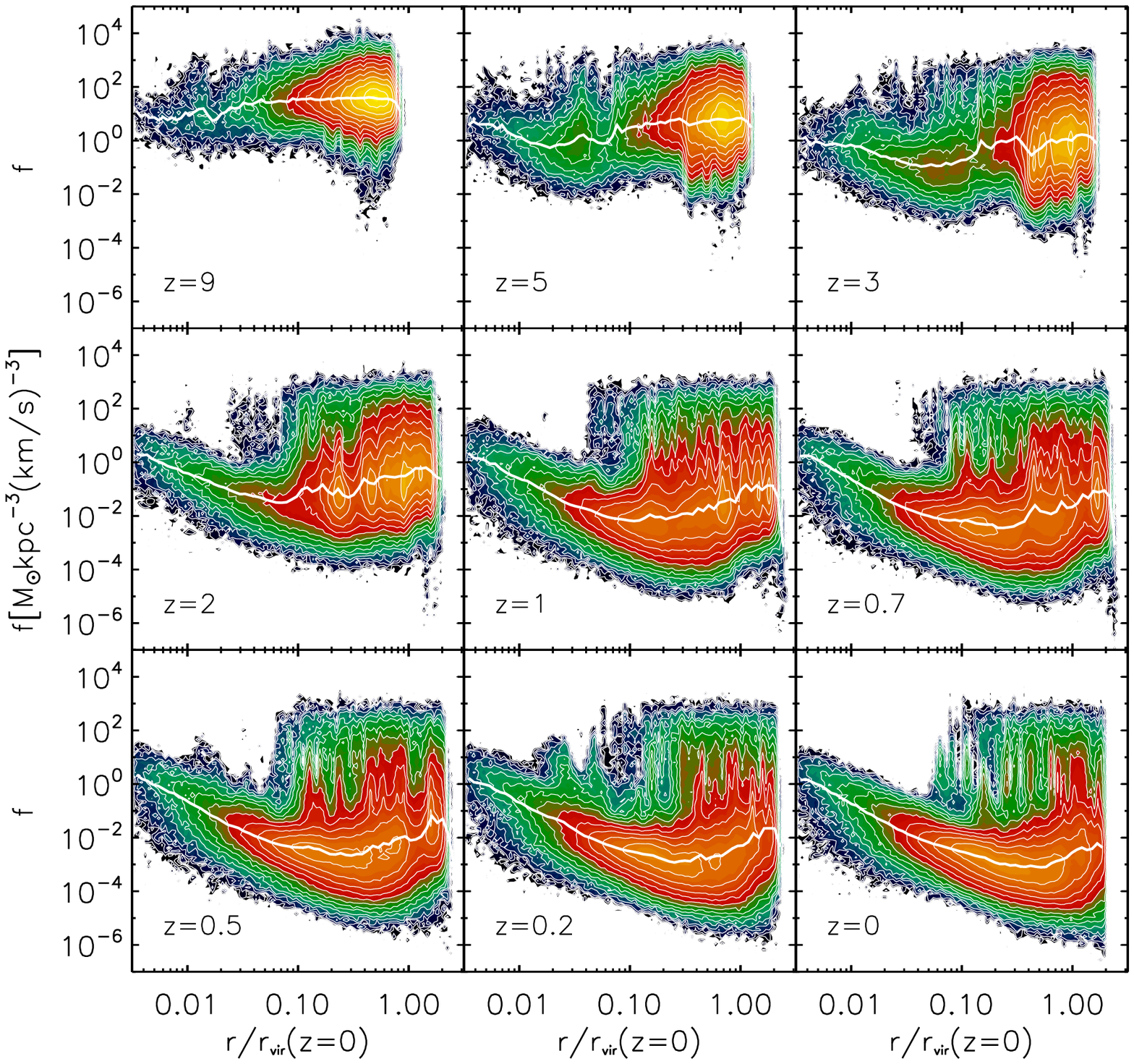,width=16.cm}}
\caption{Contours of constant particle number density in the plane
($\log(f), \log(r)$) for the $G_1$ halo, plotted at different redshifts.
At each redshift the radial distribution of particles is given
relative to the most bound halo particle in units of the virial radius
of the halo at $z=0$. The yellow and black contours correspond to the
regions with the highest and lowest particle number densities
respectively.  Contours are spaced at logarithmic intervals in
particle number density relative to the maximum density contour.  $F$,
the median value of $f$ at each radius is represented by the
solid white line.
\label{fig:rf_L25_cont}}
\end{figure*}

It is particularly illuminating to plot the full coarse-grained phase
space density of all particles as a function of radius from the center
of mass of the main halo at each redshift
(Figure~\ref{fig:rf_L25_cont}).  At any radius from the center there
exist particles with phase-space densities spanning between 4-8
decades in $f$. The colored contours represent a constant particle
number per unit area on the plane $(\log(f), \log(r))$.  The
yellow/orange contours represent the parameter range with the largest
number of particles, while the blue/black contours represent regions
with the smallest number of particles.  The solid white curves on each
plot show the median value ($F$) of $f(\mathbf{x}, \mathbf{v})$ in 100
logarithmically spaced bins in $r$. These curves are identical to the
various curves in Figure~\ref{fig:f_allz} and trace the regions with
the largest particle number density at each radius.  Numerous spikes
in $f$ are seen at $r > 0.1r_{\rm vir}$ and correspond to DM
sub-halos.

As the evolution proceeds, there is an overall lowering of the median
phase-space density $F$ (white curves).  There is also a continuous
decrease of the low-end envelope of $f$ to lower values pointing to an
increasingly large amount of matter with low phase-space
density. However the upper envelope representing the highest phase
space density regions at the centers of the subhalos remains
relatively unchanged with redshift, indicating that primordial values
of $f$ are largely preserved at the centers of DM subhalos lying
outside the center of the main halo.

\begin{figure*}
\centerline{\psfig{figure=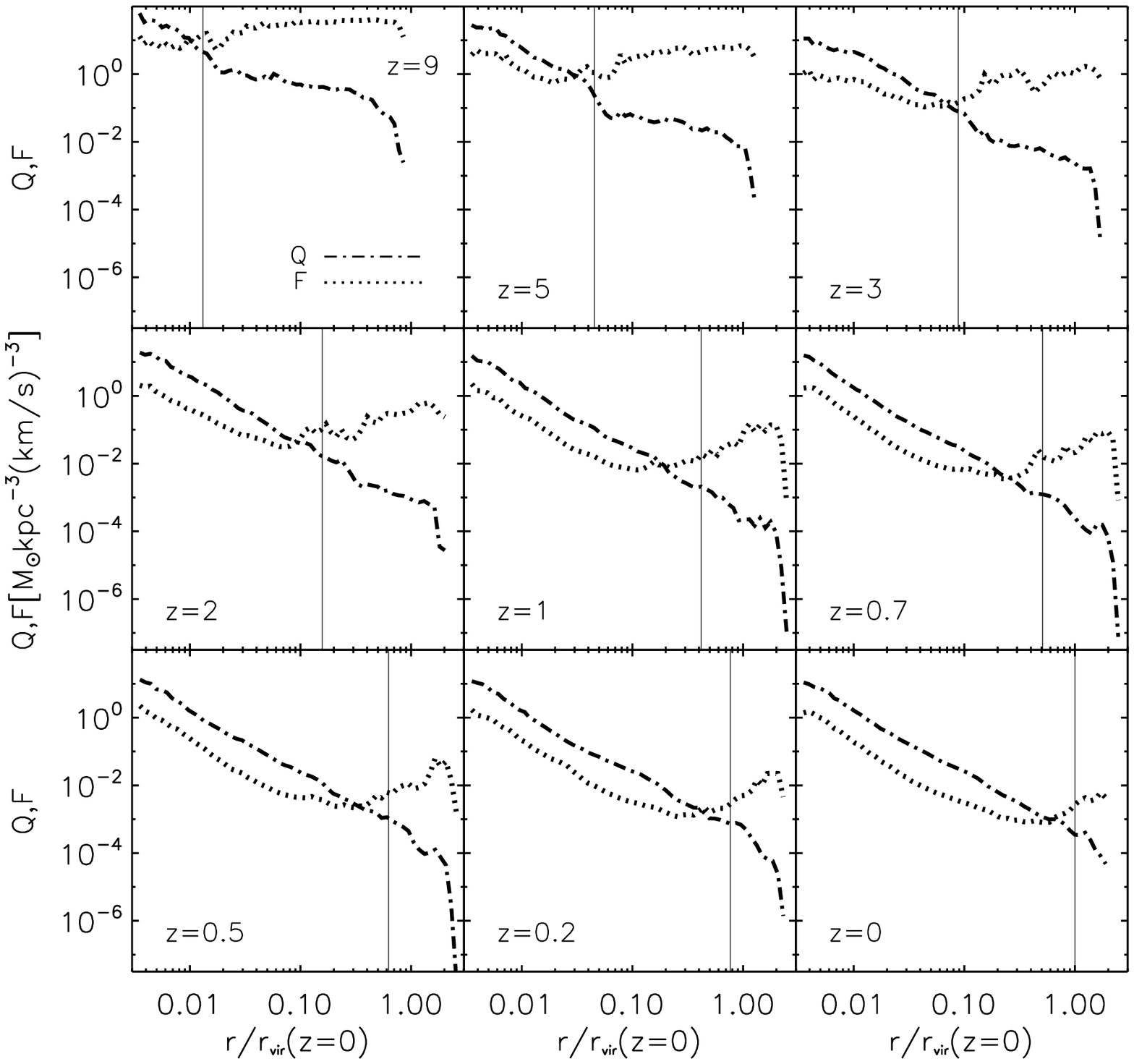,width=16.cm}}
\caption{$Q=\rho/\sigma^{3}$ (dot-dashed line) and median $F$ (dotted line) as
a function of $r/r_{\rm {vir}} (z=0)$ for different redshifts for the $G_1$ halo.
The dashed vertical line is situated at the virial radius of the
main halo at each redshift.
\label{fig:qf_L25}}
\end{figure*}

It is illustrative to compare the evolution of $F(r)$ and $Q(r)$ with
redshift.  Figure~\ref{fig:qf_L25} shows the two curves plotted as a
function of radius (in comoving units) at each redshift for all the
matter in halo $G_1$ that lies within two virial radii at $z=0$, as a
function of physical radius from the center of the most massive
progenitor of the final halo (in units of the virial radius at $z=0$).
In each panel a thin vertical line is drawn at {\it the virial radius
of the halo at that redshift}.  

We find the best fit power-law $Q \propto r^{-1.84 \pm 0.012}$ at
$z=0$ to the profile within $0.6r_{\rm vir}$. In agreement with
\citet{hoffman07}, we find that the same power-law provides a
reasonably good fit to the profiles at redshifts from $z\sim 5$ to
$z=0$.  While $Q$ always decreases with radius since it is a spatial
average over increasingly large volumes of configuration space, the
median phase-space density $F$ decreases monotonically with radius
only within about $0.6 r_{\rm vir}$ of the main progenitor at that
redshift. Outside the virial radius at each redshift (i.e. to the
right of the vertical dashed line), $F(r)$ become significantly
flatter, even increasing with increasing radius. At all $z \ne 0$ this
represents the median phase-space density of matter that will lie
within the outer radius of the halo at $z=0$, but is either still
unvirialized or lies within small halos.  In general this material is
not as mixed as the more massive main progenitor.  The nearly constant
value of $F$ beyond the virial radius at high $z$ may be interpreted
as the median phase-space density of DM in the Universe at that
redshift. At lower redshifts, a non-negligible fraction of this matter
lies within inner regions of small halos, which have undergone
relatively small amount of mixing. As more and more material undergoes
substantial mixing as the halos evolve, the high phase-space density
unmixed material in the subhalo centers, as well as in the relatively
unmixed streams formed from disrupted subhalos, prevents the median at
large radii from decreasing rapidly, despite the large increase in low
material with low $f$-values within the virialized regions of the main
halo.

\begin{figure*}
\centerline{\psfig{figure=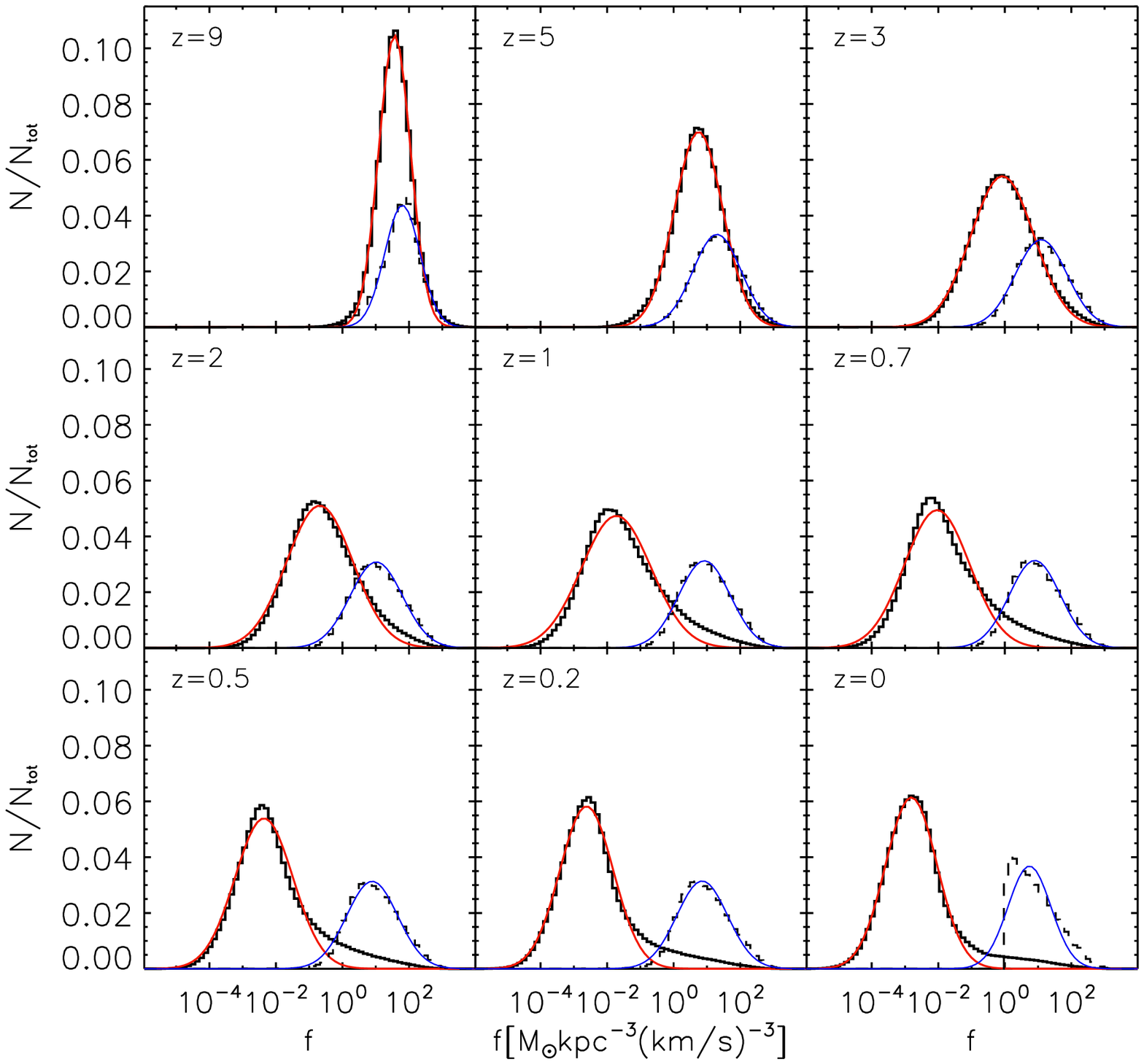,width=16.cm}}
\caption{Histograms of the phase-space density $f$ of all halo
particles as a function of redshift (thick solid histograms).  The 
dashed histograms follow those particles which have the highest
phase-space densities $f \ge 1$ at $z=0$, with ordinate values multiplied by 10 to enhance visibility.
The thin red curves are the best-fit Gaussians to all the particles in the thick solid histograms, 
while the thin blue curves are best-fit Gaussians to dashed histograms.
\label{fig:hist_L25}}
\end{figure*}

To better understand the evolution of $f$ with redshift we plot
histograms of $\log(f)$ at each redshift in Figure~\ref{fig:hist_L25}
(thick solid histograms).  The thin red curves at each redshift
represent the best-fit Gaussians to the distribution of $\log(f)$ at
that redshift. A Gaussian provides a good fit to
the majority of the mass.  (Since $\log(f)$ is close to Gaussian, $f$
itself has a log-normal distribution.)  By approximately $z = 1$, the
skewness of the distribution is significant, and the skewness
increases steadily until $z=0$. To better understand the origin of the
matter in the high-$f$ tail at z=0, we plot the distribution of DM particles that have the values of $f \ge 1\funits~$ at
$z=0$. The ordinate values of this distribution (multiplied by a
factor of 10 to enhance visibility), are shown in the dashed
histograms.  The thin blue curves show that the high-$f$ tail is also
well fitted by a Gaussian (except at $z=0$).  As we saw in
Figure~\ref{fig:rf_L25_cont}, most of these high-$f$ tail particles
are in subhalos at $z=0$, while some are also in the highest
phase-space density particles in the central cusp of the halo at
$z=0$. This high-$f$ sub-population has a Gaussian distribution at
$z=9$ with a mean $f=1.73 \pm 0.6\funits~$ compared with the mean
$f=1.57 \pm 0.54\funits$ for all the particles (in the solid curve) at
$z=9$.  A Students' T-test indicated with 99\% confidence that both
distributions are drawn from the same population at $z=9$. This
implies that the material that lies in the centers of DM subhalos at
$z=0$ will have phase-space densities that are representative of the
mean phase-space density of matter at $z=9$.


\bigskip
\section{Comparison of four Milky-Way sized halos}
\label{sec:otherhalos}

In this section we present results for the phase-space DFs for all
four of the DM halos described in \S~\ref{sec:numerical}. Although we present results
mainly at $z=0$, evolution with redshift for each of the halos were similar to the evolution of 
halo $G_1$ presented earlier.

\begin{figure}
\centerline{\psfig{figure=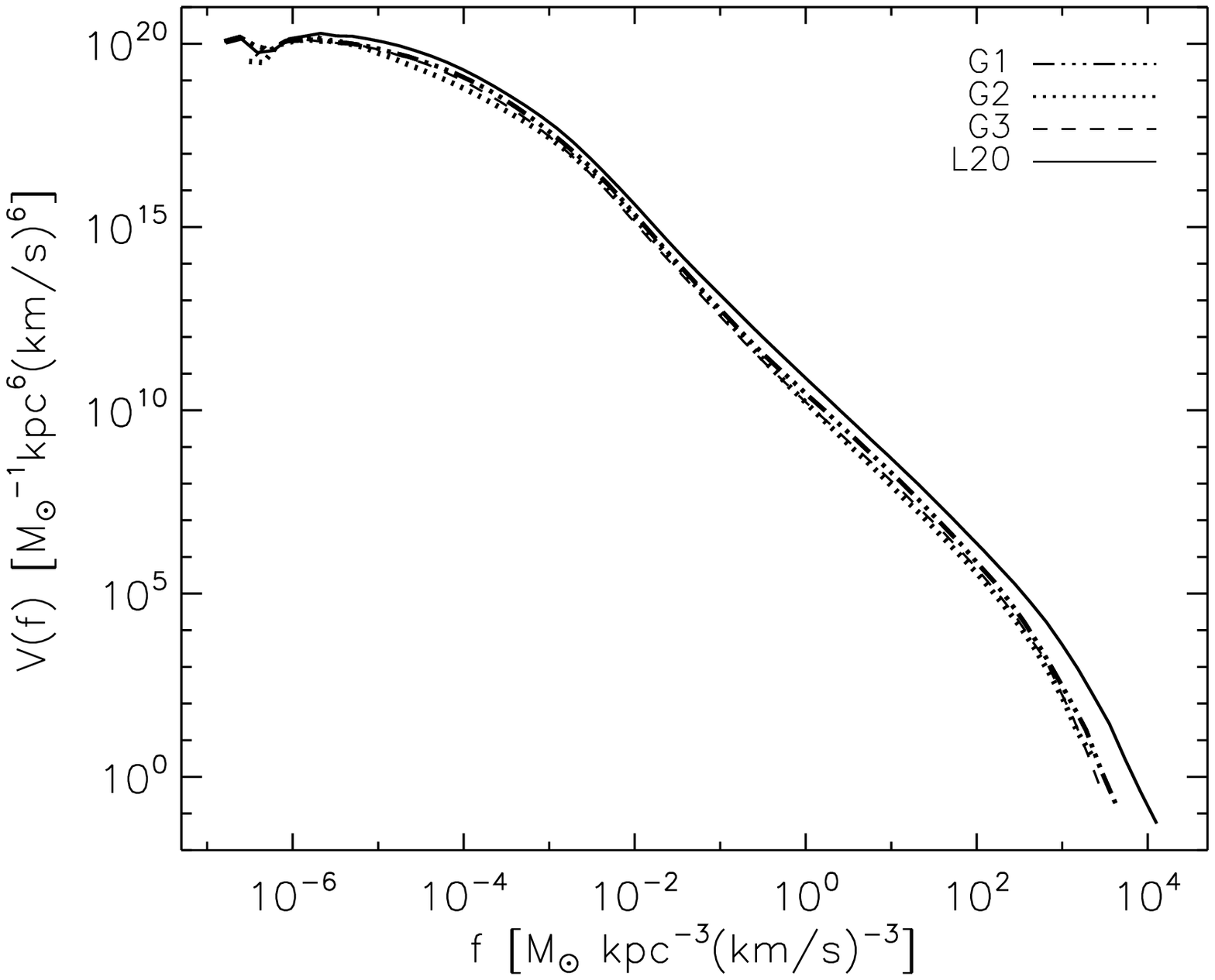,width=8.5cm}}
\caption{$V(f)$ for four halos, at $z=0$.  The triple-dot-dashed
line is for $G_1$, the dotted line for $G_2$, the dashed line for $G_3$, and the
solid line is for higher mass resolution simulation $L20$.
\label{fig:vf_halos}}
\end{figure}

Figure~\ref{fig:vf_halos} compares the volume distribution of phase
space density $V(f)$ for the four different halos at $z=0$.  All three
halos from the $L25$ simulation ($G_1$, $G_2$, $G_3$) show almost identical
profiles in $V(f)$ confirming the universality of the process that
produced the phase-space DF. For halo $L20$, $V(f)$ lies systematically
above the other curves especially at higher values of $f$, where it
also extends to large values of $f$. It was first shown by \citet{shar_stein06} 
that this is a numerical consequence of
the increased mass resolution of the simulation. This indicates that the 
absolute value of $f$
derived in the previous section is somewhat dependent on the mass
resolution of the simulation and increases slightly with increasing
mass resolution.  In all four halos $V(f)$ is well approximated by a
power-law over nearly six orders of magnitude in $f$. The first column
of Table~\ref{tab:halos_fits} gives details of the power-law fits to
$V(f)$ (i.e. the power-law slopes and their errors over the range
$10^{-4} < f < 10^{2.5}$.) The power-law slopes we obtained are
similar to those obtained by previous authors \citep{Arad04}.

\begin{figure}
\centerline{\psfig{figure=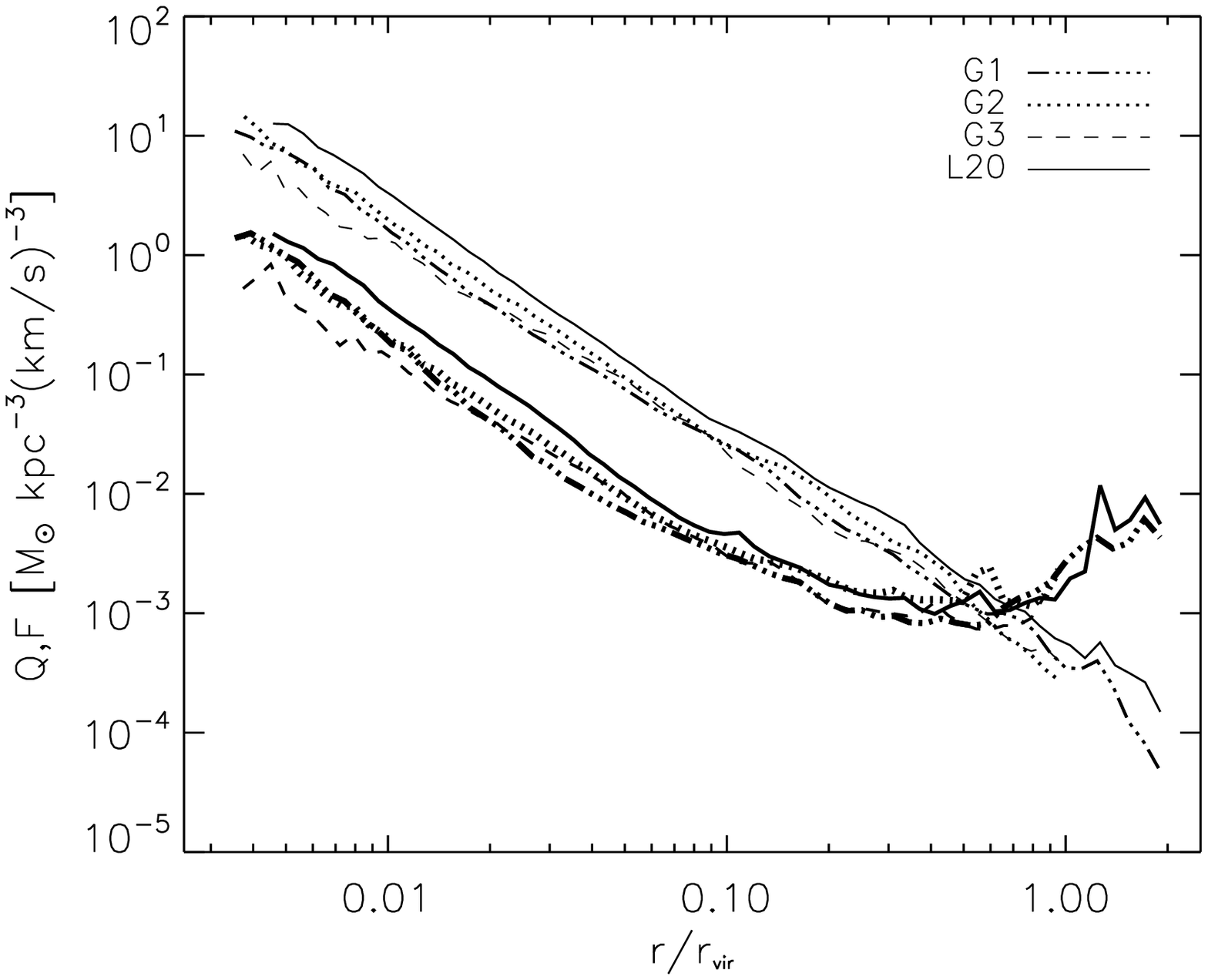,width=8.5cm}}
\caption{$Q$ (thin lines) and $F$ (thick lines) for all the halos studied, at
$z=0$.  Triple-dot-dashed lines are for $G_1$, dotted lines for $G_2$, dashed
lines for $G_3$, and the solid lines are for $L20$.
\label{fig:qf_halos}}
\end{figure}

In Figure~\ref{fig:qf_halos} we plot $F(r)$ and $Q(r)$ for the four
different halos at $z=0$. The top set of four curves show $Q(r)$ while
the lower set of curves show $F(r)$.  In all four halos, $Q(r)$ is
well fitted by a power-law of slope $\beta = 1.8 - 1.9$ (see
Table~\ref{tab:halos_fits}).  $F(r)$ shows significant deviations from
a simple power-law profile, with a systematic upturn beyond $r/r_{\rm
vir} > 0.1$.  The higher resolution simulation ($L20$) appears to have
systematically higher $F$ and $Q$ values than the halos from the low
resolution simulation  \citep[again, a numerical consequence of the higher mass
resolution][]{shar_stein06}. Table~\ref{tab:halos_fits} gives
the values for the slopes and the error-bars on the power-law fits to
$Q(r)$ and $F(r)$ for $r < 0.6 r_{\rm vir}$ as well as the inner
power-law slope of $F(r)$.

\begin{table*}
\centering
\caption{Power-law indices for fits to $V$, $Q$, and $F$ for four halos at
$z=0$ for $r < 0.6r_{\rm vir}$ }\label{tab:halos_fits}
\begin{tabular}{@{}lccc}
\hline
Halo & $V$ & $Q$ & $F$ \\
\hline
G$_{1}$ &$-2.34 \pm 0.02$  & $-1.84 \pm 0.01$ & $-1.59 \pm 0.05$  \\
G$_{2}$ &$-2.37 \pm 0.02$  & $-1.82 \pm 0.01$ & $-1.46 \pm 0.06$  \\
G$_{3}$ &$-2.35 \pm 0.02$  & $-1.75 \pm 0.01$ & $-1.42 \pm 0.03$  \\
L20         &$-2.27 \pm 0.01$  & $-1.87 \pm 0.01$ & $-1.64 \pm 0.07$  \\
\hline
\end{tabular}
\end{table*}

\bigskip
\section{Discussion}
\label{sec:origin}

Several previous studies have attempted to account for the origin of
the power-law $Q(r)$ profiles of DM halos.  Notably, it has been
argued that power-law profiles result from virialization and not from
the hierarchical sequence of mergers, since they are also produced in
simple spherical gravitational collapse simulations
\citep{taylorN01,barnes06, barnes07}. Our results presented in the
previous section (Figure~\ref{fig:qf_L25}) show that while $Q(r)$ and
$F(r)$ have approximately power-law form within $0.6 r_{\rm vir}$ at a
given redshift, $F(r)$ flattens out and remains quite flat beyond this
radius. Furthermore, as the hierarchical growth of the halo progresses
these approximately power-law profiles extend to larger radii until
they encompass all the mass within the virial radius at $z=0$. 

It has been shown from both theoretical arguments and numerical
simulations \citep{dehnen2005,vass_etal_08a} that power-law profiles
of phase-space density for central cusps are well preserved during
major mergers. At intermediate times during a major merger, deviations 
from the power-law profile are seen but these largely disappear at the 
end of the merger. In major mergers there are two main reasons for the
preservation of the power-law profiles. First, the most tightly bound
material - that forming a steep central cusp or shallow core preserves
its phase-space density in the final remnant. This is a consequence of
the additivity of the excess mass function - a result of the fact that
steeper cusps are less mixed than shallower cusps \citep{dehnen2005}.
Over 60\% of the material in the central cusp (within one scale radius) of a
progenitor NFW halo remains within the cusp of the merger remnant
\citep{valluri07}.  The second reason for preservation of power-law
profiles at large radii is that material outside three scale radii of the
progenitor halos is redistributed to other radial bins almost
uniformly from each radial interval. In addition, nearly 40\% of the
material within the virial radius of the progenitor halos is ejected
to beyond the virial radius of the final remnant
\citep{kazantzidis_etal06, valluri07}.  This material can be shown
to have originated in roughly equal fractions from each radial
interval beyond three scale radii and has higher phase-space density than
expected from the simple power-law extrapolation of the inner
power-law. This self-similar redistribution of material
contributes to the preservation of power-law profiles in
$Q(r)$ and $F(r)$ in equal mass binary mergers. Thus, once a power-law
phase-space DF has been established in a DM halo, major mergers will
not destroy this profile. This has been confirmed by our recent work on mergers of equal mass
halos \citep{vass_etal_08a}.

As discussed in the previous section (Figure~\ref{fig:hist_L25}), the
coarse-grained phase-space density $f$ in $\Lambda$CDM halos has a
nearly log-normal distribution with the median of $\log(f)$ (the peak of the
histogram $f_{\rm peak}$) evolving steadily toward lower values of $f$
as the halo grows.  We now quantify the evolution of $f_{\rm peak}$ with redshift.

\begin{figure}
\centerline{\psfig{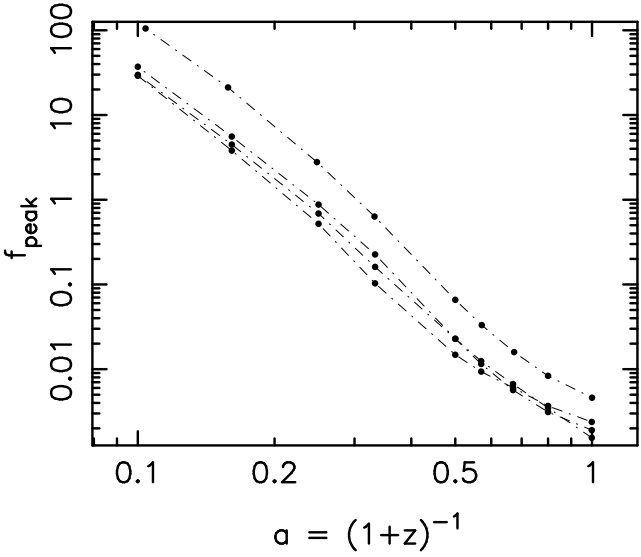}}
\caption{The median phase-space density (peak of histograms $f_{\rm peak}$ in
Fig.~\ref{fig:hist_L25}) at each redshift as a function of the cosmic  scale factor $a= (1+z)^{-1}$, plotted for four Milky-way
sized halos (points). The thin dotted lines connect points for the four
individual halos.  
\label{fig:medf}}
\end{figure}

Figure~\ref{fig:medf} shows the evolution of median phase-space
density $f_{\rm peak}$ derived from the histograms of $\log(f)$ in
Figure~\ref{fig:hist_L25} for each of the 4 Milky-Way sized
cosmological halos in this study.  The solid dots represent the values
of $f_{\rm peak}$ as a function $a$, while the dotted curves are meant
to guide the eye by connecting points for each of the 4 individual
halos.  For $a<0.5$ the decline in $f_{\rm peak}$ is approximately 
power-law: $f_{\rm peak}(a) \propto a^{-4.5}$. 
 This implies that the matter inside the halos
has undergone significant mixing due to virialization during the
hierarchical formation process, the leveling off of the curve at $a\sim 0.5$ indicates a 
decrease in mixing after $z=1$ , possibly resulting from a decrease 
in mass accreted in major mergers.

On average, as a halo grows via accretion its median density
decreases as a power-law with time, despite the fact that the most
centrally concentrated material retains its original high phase-space
density. This nearly power-law profile in $f_{\rm peak}$ leads to some
insights into the development of phase-space density profiles.  We can
break down the formation of halos into two phases \citep{li_etal07}:
the fast accretion regime during which halo mass grows very rapidly
and the slow accretion regime.  In the fast accretion regime, the
mixing processes are very efficient as the potential well is
established and potential fluctuates rapidly and constantly. This
results in a rapid decrease in the overall central phase-space
density of the halo (as seen from the rapid drop in the central most
regions of the profiles in Fig.~\ref{fig:f_allz}.)  The inner profile
of phase-space distribution should be largely set during this stage.
As halos grow subsequently, either by major mergers or quiescent
accretion of smaller halos, the average phase-space density decreases,
but the pre-existing high central phase-space density cusps are 
preserved.

At these later times the evolution of the true phase-space density is
complex and occurs due to the accretion of high phase-space density
material in subhalos, as well as loosely bound material at the
edges of the halo. The steady decrease in the amplitude and increase
in the slope of the $F(r)$ profiles in Fig.~\ref{fig:f_allz} with
redshift show that the true phase-space density does not obey the
power-law profiles seen in $Q(r)$ at all redshifts.  Variations in the
individual power-law profiles of halos both at $z=0$
(Fig.~\ref{fig:qf_halos}) and with redshift reflect the large
variation in the cosmic accretion histories of individual
halos. Additional deviations from the power-law arise due to the
matter bound to subhalos that survives with high phase-space density
and leads to the variation of the coarse-grained $f$ of some six
orders of magnitude in the outer regions of halos.

All this indicates that the actual phase-space distribution is not as
universal and simple as the $Q(r)$ profiles lead one to believe.  The
mechanism behind the universal power-law form of the $Q(r)$ profiles
is therefore likely to be different (e.g., it is not affected by the
presence of subhalos) and simpler than the processes that shape the
distribution of the coarse-grained phase-space density.  The $Q$
profile is a ratio of the density and velocity dispersion (which can
be interpreted as a measure of temperature) and is therefore related
to the entropy. For an ideal monatomic gas the entropy can be defined
as $K_{\rm gas}=T/\rho^{2/3}$, while for DM, by analogy, the
entropy can be defined as $K_{\rm dm}=\sigma^2/\rho_{\rm dm}^{2/3}$
\citep{faltenbacher_etal07}.  This, as noted by \citet{hoffman07},
gives $K_{\rm dm}\propto Q^{-2/3}$ or $K_{\rm dm}\propto r^{1.2}$ for
$Q\propto r^{-1.8}$.  This power-law form and slope of the entropy
profile is very similar\footnote{The slope is even more similar if one
takes into account the random bulk motions of the gas in estimating
$K_{\rm gas}$ \citep{faltenbacher_etal07}.} to the one found for the
gas in the outer regions of clusters in cosmological simulations
\citep[e.g.,][]{borgani_etal04,voit_etal05}.  This power-law is also
predicted by the models of spherical accretion \citep{tozzi_norman01}
and reflects the increasing entropy to which the accreting material is
heated as the halo grows its mass. Although the processes governing
the virialization of gas and DM are different (the
short-range local interactions for the former, and long-scale
interactions for the latter), the fact that the resulting entropy
profiles are quite similar indicate that they lead to the same
distribution of entropy. The $Q(r)$ profile therefore may reflect the
overall entropy profile of DM, not the coarse-grained local
phase-space density, which exhibits a more complicated behavior.

\bigskip
\section{Summary and Conclusions}
\label{sec:conclude}

We have investigated the evolution of the phase-space density of the
DM in cosmological simulations of the formation of Milky-Way-
sized DM halos. The analysis was carried out using two
different codes for estimating the phase-space density. Both codes
give qualitatively similar results, but the estimated values of phase-
space density $f$ are quite sensitive to the type of code and, for a
given code, also depend quite sensitively on the choice of smoothing
kernel used. Based on comparisons of the two codes (\fiestas and
\enbid) and various smoothing parameters (Appendix), we select the
\enbid code with $n=10$ kernel smoothing and present results for the
analysis with this set of parameters.

The simulations presented in this paper complement our analysis of the
evolution of phase-space density in binary major mergers
\citep{vass_etal_08a}.  We confirm that the profiles of
$Q(r)=\rho_{\rm dm}/\sigma^3_{\rm dm}$ computed by previous authors
can be described by a power-law $Q(r)\propto r^{-1.8\pm 0.1}$ over
more than two orders of magnitude in radius in all halos.  The median
of the phase-space density ($F(r)$) at given radius $r$, however,
exhibits a more complicated behavior. Although $F(r)$ is approximately
a power-law for $r< 0.6r_{\rm vir}$, the profiles generally flatten in
the outer regions. Subhalos contribute somewhat to this behavior,
although their effect is limited by the relatively small fraction of
mass ($<0.1$) bound to them. However, in addition to subhalos, a
significant fraction of high phase-space density matter is in the
relatively unmixed material (possibly in streams) arising from tidally
disrupted subhalos \citep[e.g.,][]{Arad04,diemand_etal08}. The rise in
$F(r)$ at large radii suggests that the fraction of mass in such
streams can be substantial.

This behavior holds at earlier epochs.  From $z=5$ to $z=0$, material
within $r_{\rm {vir}}(z)$ at each redshift follows a power-law in $Q$
with an approximate power-law slope of $\sim -1.8$ to $-1.9$. In
contrast, $F(r)$ can only be well described by a power-law in the inner
regions, and its slope changes continuously with redshift. Beyond the
virial radius, $Q$ (a quantity that is obtained by averaging over
increasingly large volumes) decreases rapidly with radius, but the
median value of $F$ flattens significantly.  We argue that $F$ is a
more physically meaningful quantity, especially for understanding the
evolution of phase-space density in collisionless DM halos, as it
measures the median of the true coarse-grained phase-space density.

At all redshifts, the highest values of phase-space density $f$ are
found at the centers of DM subhalos.  In the center of the
main halo, the median phase-space density ($F$) drops by about an
order of magnitude from $z=9$ to $z=0$.  In contrast, the centers of
DM subhalos maintain their high values of $f \sim 10^3\funits$ at all
redshifts. The highest values of $f$ at the center of the main halo are,
therefore, lower and less representative of the primordial phase-space
density of DM particles than the central value of $f$ in the
high phase-space density subhalos. At $r_{\rm {vir}}$, the decrease in
median phase-space density is much more significant, with $F \approx
30\funits$ at $z=9$ decreasing to $ F\approx 10^{-3}\funits$ at
$z=0$, a decrease of over four orders of magnitude
(Figure~\ref{fig:rf_L25_cont}).

The evolution of $F(r)$ and $V(f)$ with redshift are consistent
with expectations from the Mixing Theorems, which require that mixing
reduces the overall phase-space density of matter in collisionless
systems and that the volume of phase-space associated with any value
of $f$ increases due to mixing and relaxation.
 
The distribution of $f$ is approximately log-normal until $z\sim
3$. As time progresses, the mean and median of $\log(f)$ shift to
progressively lower values as a larger and larger fraction of matter
undergoes mixing and moves to lower values of $f$.  Some fraction of
high phase-space density material does survive in the centers of
subhalos and in the relatively unmixed streams leftover after subhalo
disruptions, which skews the distribution.  Remarkably, the highest
phase-space density particles at $z=0$ have retained their phase-space
density since $z \approx 9$, the earliest epoch we analyzed.  The
phase-space density in the centers of DM subhalos is
therefore representative of the mean phase-space density of DM at the
highest redshifts. This can potentially allow for stronger constraints
to be placed on the nature of DM particles from the Tremaine-Gunn bound
\citep{tremaine_gunn_79, hogan_dalcanton_00}.

The median value of phase-space density decreases with decreasing
redshift approximately as a power-law described by $f_{\rm peak}
\simeq a^{-4.5}$. This majority of the decrease in $f_{\rm peak}$,
is the result of mixing within virialized halos which reduces
the coarse-grained phase-space density of matter that has turned
around from the Hubble flow, much of this mixing occurs prior to $z\sim 1$, 
after which the rate of mixing in galactic sized halos slows down.

\bigskip

\section*{Acknowledgments}

We thank Y. Ascasibar and S. Sharma for the use of the \fiestas code
and \enbid code, respectively.  We especially thank S. Sharma for
detailed discussions on his \enbid code. We thank the referee Stephane 
Colombi for his helpful comments. IMV would like to thank
Stephen T. Gottesman for valuable input during the early stages of
this work, and acknowledges the support of National Science Foundation
(NSF) grant AST-0307351 to the University of Florida.  IMV, MV, AVK,
and SK were supported in part by Kavli Institute for Cosmological
Physics at the University of Chicago (where most of this work was
carried out) through grants NSF PHY-0114422 and NSF PHY-0551142 and an
endowment from the Kavli Foundation and its founder Fred Kavli.  
MV is also supported by the the University of Michigan. AVK
is also supported by the NSF under grants AST-0239759 and AST-0507666
and by NASA through grant NAG5-13274. AVK would like to thank the
Kavli Institute for Theoretical Physics, supported in part by the NSF
under grant PHY05-51164, for hospitality during the final editing of
the paper and participants of the workshop ``Back to the Galaxy'' for
useful discussions and feedback on the results of this paper.  SK is
supported by the Center for Cosmology and Astro-Particle Physics
(CCAPP) at The Ohio State University.

\bigskip

\appendix
\section{Comparison of  ``\fiestas''  and ``\enbid'' Analysis of Halo $G_1$}

The numerical estimation of coarse-grained phase-space densities
during the evolution of the four N-body halos presented in this paper was
carried out using two publicly available codes ``\fiestas''
\citep{ascasibar_binney05} and ``\enbid'' \citep{shar_stein06}. A
comparison of these two codes has previously been presented by
\citet{shar_stein06}, who showed for an analytic DF (for the
spherical Hernquist potential), that ``\enbid with $n=10$ kernel
smoothing'' gave the highest fidelity to the analytical DF.  In a
related work \citet{vass_etal_08a} confirmed the findings of
\citet{shar_stein06} for a spherical isotropic NFW halo. This latter
study is the main basis for our choice of \enbid with $n=10$ kernel
smoothing.  After this paper was submitted to the Journal we became aware of
the work of \citet{maciejewski_etal_08}. These authors carried out a similar, 
and somewhat more detailed comparison, of coarse-grained  phase-space 
density estimators for N-body simulations on cosmological halos at $z=0$. Our results are in agreement with theirs.

\begin{figure}
\centerline{\psfig{figure=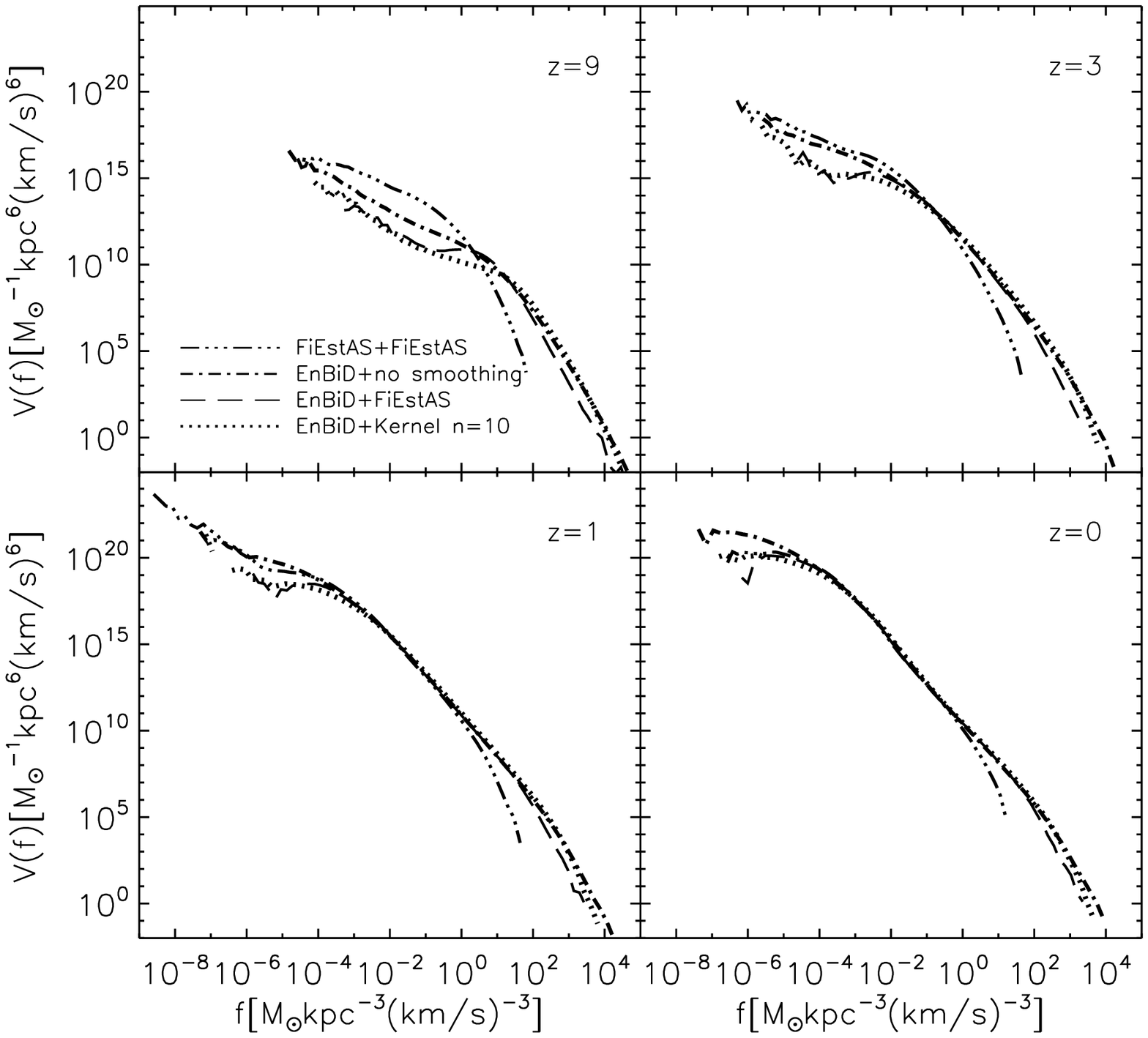,width=8.5cm}}
\caption{The volume DF $V(f)$, at different redshifts during the
evolution of the $G_1$ halo in the $L25$  simulation.  Plots compare results obtained with the \fiestas code and \enbid code (using three different parameters) as indicated in the line
legends.
\label{fig:fvfz}}
\end{figure}

However, it is unclear whether the comparisons with analytic profiles
of isolated halos at $z=0$ are valid for matter distributions arising
from cosmological N-body distributions, at high redshifts,
where the majority of the particles actually lie outside virialized
halos. Our purpose in this Appendix is to present a comparison of
results obtained with the different codes at a range of redshifts to
allow readers to appreciate how sensitive some of the results
presented in this paper are to the choice of code and smoothing
parameters used in density estimation. 
Since the coarse-grained DF $f(\mathbf{x}, \mathbf{v})$ is a
6-dimensional function, it is difficult to compare estimates for this
function obtained using different codes. Traditionally the best single
variable function to compare is the volume DF $V(f)$.  Variations in
the estimation of $f$ translate to variations in estimation of $V(f)$.
In Figure~\ref{fig:fvfz} we compare the volume density of phase-space
$V(f)$ obtained for the $G_1$ halo at four different redshifts using the
\fiestas code (triple-dot-dashed curves), \enbid code with no
smoothing (dot-dashed curves), \enbid with \fiestas smoothing
(long-dashed curves), and \enbid with a $n=10$ kernel (dotted
curves). In each case we see that $V(f)$ has a nearly power-law
distribution over more than six orders of magnitude in $f$ at $z=0$
(from $\sim 10^{-4}$--$10^{3}\funits$). Table~\ref{tab:voffslopes}
gives the slopes of power-law fits for the four different estimates of
$V(f)$ over the range $10^{-4} < f < 10^{3}$ at $z=0$. The \fiestas
estimate of $V(f)$ is systematically lower than all \enbid estimates
at high values of $f$ (at all redshifts).  In addition all \enbid
curves extend to much higher values of $f$ than the \fiestas estimate
(this is particularly true at $z=9$ where the \fiestas estimate
differs from the other estimates both quantitatively and
qualitatively).  The results from the various \enbid estimates differ
very little at intermediate values of $f$ ($10^{-4}$--$10^{3}$) 
and consequently we are confident that conclusions drawn from
the median $f$ is quite insensitive to the details of the \enbid
parameters used to obtain $f$.

\begin{table}
\centering
\caption{\label{tab:voffslopes} Power-law indices $V(f) \propto f^{-\alpha}$ and $F(r)\propto r^{-\beta}$ at $z=0$ from different codes}
\begin{tabular}{lcc}
\hline
Code                                      &  $\alpha$                & $\beta$       \\
\hline
\fiestas                                & $2.62 \pm 0.06$   & $1.43 \pm 0.02$\\
\enbid  (no smoothing)        & $2.35 \pm 0.02$  & $1.65 \pm 0.02$\\
\enbid (\fiestas  smoothing) & $2.46 \pm 0.03$  & $1.69 \pm 0.07$\\
\enbid (kernel $n=10$)        & $2.34 \pm 0.02$  & $1.59 \pm 0.05$\\
\hline
\end{tabular}
\end{table}

\begin{figure}
\centerline{\psfig{figure=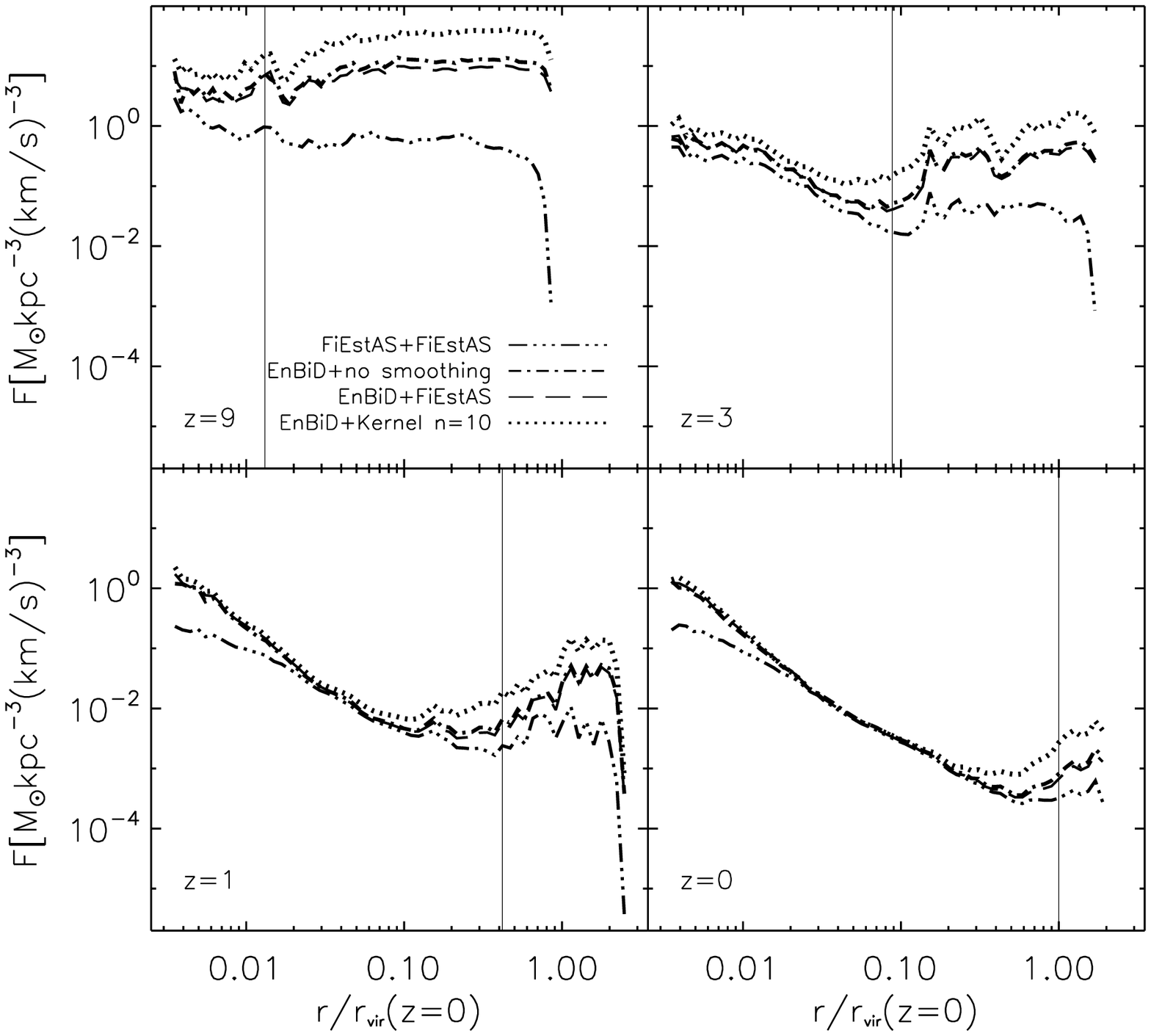,width=8.5cm}}
\caption{ $F$ for the $G_1$ halo in the $L25$ simulation at different
redshifts obtained using \fiestas and \enbid as indicated by
line-legends.  The dashed vertical line is situated at the virial
radius of the main halo at each redshift.
\label{fig:allfzs}}
\end{figure}

The differences between the various estimates of $F$ at large radii become
significantly larger at higher redshift. In Figure~\ref{fig:allfzs} we plot the
four different estimates of $F(r)$ at four different redshifts in the evolution. 
The vertical dashed line in each panel represents the virial radius of the main
halo at that redshift. We see that $F$ from any of the codes is quite flat
beyond the virial radius in all cases, but the absolute values of the curves
differ significantly. Note that at $z=9$, the different estimates can differ by
as much as two orders of magnitude. 

In this paper we choose to present results from \enbid with $n=10$
kernel smoothing over the other estimates largely because it does the
best job of reproducing analytic DFs \citep{shar_stein06} and because
it appears to provide a good upper limit to the phase-space density
both at high and low values of $f$. In the absence of an analytic 
comparison of the estimates at high redshift, we caution the reader to refrain
from drawing very strong conclusions regarding absolute values of $f$
or $F$ from the results presented here.

\bibliographystyle{mn2e}
\bibliography{main2}

\bsp

\label{lastpage}

\end{document}